\documentstyle[aps,prb]{revtex}
\begin{document}

\title{Thermal Conductivity near H$_{c2}$ for spin-triplet
superconducting States with Line Nodes in Sr$_2$RuO$_4$}
\author{L. Tewordt and D. Fay}
\address{I. Institut f\"ur Theoretische Physik,
Universit\"at Hamburg, Jungiusstr. 9, 20355 Hamburg, 
Germany}
\date{\today}
\maketitle
\begin{abstract}
        We calculate the thermal conductivity $\kappa$ in magnetic fields near
H$_{c2}$ for spin-triplet superconducting states with line nodes vertical and
horizontal relative to the RuO$_2$-planes. The method for calculating the 
Green's functions takes into account the spatial variation of the order
parameter and superconducting flow for the Abrikosov vortex lattice. For 
in-plane magnetic field we obtain variations of the in-plane $\kappa$ with 
two-fold symmetry as a function of rotation angle where the minima and 
maxima occur for field directions parallel and perpendicular to the heat flow.
The amplitude of the variation decreases with increasing impurity  scattering
and temperature. At higher temperatures the minima and maxima of the 
variation are interchanged. Since the results for vertical and horizontal line 
nodes are almost the same we cannot say which of the two pairing models is 
more compatible with recent measurements of $\kappa$ in Sr$_2$RuO$_4$. 
The observed four-fold modulation of 
$\kappa$ in YBa$_2$Cu$_3$O$_{7-\delta}$ is obtained for d-wave pairing by 
taking into account the particular shape of the Fermi surface and the finite 
temperature effect. The results for $\kappa$ for the f-wave pairing state with 
horizontal line nodes disagree in some respects with the measurements 
on UPt$_3$.
\end{abstract}
\vspace{0.5in}
\section{INTRODUCTION}

	The nature of the unconventional superconducting state in the layered
ruthenate Sr$_2$RuO$_4$ \cite{Maeno} is currently of great interest. The
original proposal of a spin-triplet p-wave pairing state with broken time-reversal
symmetry and a constant gap  \cite{Rice} has to be modified because recent 
specific heat \cite{Nishi} and NMR relaxation rate \cite{Ishida} experiments
indicate the presence of line nodes in the superconducting gap. These results
have lead to the proposal of new spin-triplet pairing states including f-wave
pairing states with line nodes vertical and horizontal with respect to the 
2D-planes. \cite{Hasegawa} A powerful tool for probing the anisotropic gap
structure is the thermal conductivity $\kappa$ in a magnetic field {\bf H}
because $\kappa$ is sensitive to the relative orientations of the heat current, 
the field, and the nodal directions of the gap. In fact, a modulation of $\kappa$
of 4-fold symmetry with rotation of an in-plane magnetic field reflects the
angular positions of the vertical line nodes for d$_{x^2-y^2}\:$-wave pairing in
YBa$_2$Cu$_3$O$_{7-\delta}$ (YBCO). \cite{Aubin} The observed modulation 
of the heat current in UPt$_3$ with rotation of the in-plane magnetic field 
 \cite{Suderow} seems to be in accordance with a spin-triplet f-wave pairing
state that has horizontal line nodes. \cite{WonMaki}

Recently, the thermal conductivity $\kappa$ in  Sr$_2$RuO$_4$ has been
measured in magnetic fields up to H$_{c2}$ for field directions perpendicular
and parallel to the ab-plane. \cite{Izawa,Tanatar} For rotating in-plane field the 
in-plane $\kappa$ exhibits small modulations of 2-fold and 4-fold symmetry 
with respect to the rotation angle $\alpha$. Since the amplitudes of these 
variations are much smaller than the calculated amplitudes for f-wave pairing 
states with vertical line nodes, \cite{Dahm} it has been suggested that the 
actual pairing state in Sr$_2$RuO$_4$ has horizontal line nodes like those 
proposed in Ref.~\onlinecite{Hasegawa}. Most theories of thermal conductivity 
in low fields \cite{Kubert} are based on the Doppler shift of the quasiparticle 
spectrum due to the circulating flow in an isolated vortex line. \cite{Volovik} 
Here we shall employ a quite different approach valid for fields near 
H$_{c2}$ in which both the effects of the supercurrent flow and the scattering 
of the quasiparticles by the spatial variation of the order parameter in the 
Abrikosov vortex lattice are taken into account. \cite{Brandt}

In Section II we present the theory and in Section III we discuss the results. The
conclusions are contained in Section IV.%
\vspace{0.2in}

\section{ THEORY OF GREEN'S FUNCTIONS AND THERMAL CONDUCTIVITY
IN HIGH MAGNETIC FIELDS}

     In the method of Ref.~\onlinecite{Brandt} the Gorkov integral equations for the
normal and anomalous Green's functions G and F with kernels given by the 
Abrikosov vortex lattice function are solved by expanding all functions in 
Fourier series ${\bf k}$ with respect to the sum of the spatial positions and in
Fourier integrals ${\bf p}$ with respect to the difference of the spatial
coordinates. In calculating the spatial averages it suffices to consider the
${\bf k}=0$ Fourier component. The corresponding Green's functions are given
by: \cite{Brandt}
\begin{equation}
G({\bf p},\omega)=\left[\widetilde{\omega} - \xi + 
i\sqrt{\pi} \Delta^2 |f({\bf {p}})|^2 (\Lambda/\mbox{v}\sin\theta) w(z)\right]^{-1}\,;
\label{DefG}
\end{equation}
\begin{equation}
F({\bf p},\omega)=-i\sqrt{\pi} \Delta f({\bf {p}})
(\Lambda/\mbox{v}\sin\theta) w(z) G({\bf p},\omega) \,;
\label{DefF}
\end{equation}
\begin{equation}
\Lambda = (2e\mbox{H})^{-1/2}\,; \quad \Delta^2 = \overline{|\Delta(\bf {r})|^2}\, ;
\label{Deflamdel}
\end{equation}
\begin{equation}
w(z)=\exp(-z^2) {\mathrm{erfc}}(-iz) \,;
\quad z=(\widetilde{\omega} + \xi) \Lambda/v\sin\theta \, ;
\label{Defw}
\end{equation}
\begin{equation}
\widetilde{\omega}=\omega + i \gamma \,;
\quad \gamma = \Gamma \left[ N(\omega)/N_{0}\right]^{-1} \, .
\label{Defomega}
\end{equation}
Here the applied field {\bf H} is along the c-axis, $\theta$ is the polar angle
between {\bf p} and {\bf H}, v is the Fermi velocity in the ab-plane, and 
$\Gamma$ is the normal-state impurity scattering rate. The spatial average 
$\Delta^2$ of the absolute square of the Abrikosov vortex lattice function is
approximately given for large $\kappa$  by the relations \cite{Brandt}
\begin{equation}
(\Delta\Lambda/\mbox{v})^2 =  
(\mbox{H}_{c2} - \mbox{H})/6\beta_{A} \mbox{H} \,;
\quad (\Lambda/\mbox{v})^2 = 
(\mbox{H}_{c2}/\mbox{H})\left[6\Delta_{0}^2 \right]^{-1} \, .
\label{Defdellam}
\end{equation}
Here,  $\beta_A$ = $<|\Delta|^4>/(<|\Delta|^2>)^2$ is the Abrikosov parameter 
and $\Delta_0$ is the BCS gap parameter. For $\theta \rightarrow 0$ the 
argument $z$ in Eq. (\ref{Defw}) becomes very large and 
$w(z)\sim i/\sqrt{\pi} z$ which 
yields according to Eqs. (\ref{DefG}) and (\ref{DefF}) the ordinary Green's 
functions G and F for a gap $\Delta f({\bf p})$ where $f({\bf p})$ gives 
the {\bf p}-dependence. It should be pointed out that the self-energy term 
proportional to $i w(z)$ in Eq. (\ref{DefG}) yields an imaginary part that 
corresponds to quasiparticle scattering by the vortex cores. For a field {\bf H} 
in the ab-plane with angle $\alpha$ between {\bf H} and the a-axis one has to 
make the following replacements in Eqs. (\ref{DefG}) - (\ref{Defomega}):
\begin{equation}
\sin\theta \longrightarrow \left[\sin^{2}\theta \sin^{2}(\phi-\alpha)
+ (\mbox{v}^{\prime} / \mbox{v})^2 \cos^{2}\theta \right]^{1/2} \,;
\quad (\alpha = \angle ({\bf H},{\bf a})\,) \, .
\label{sin}
\end{equation}
Here $\phi$ is the azimuthal angle of {\bf p} in the ab-plane and v$^{\prime}$ is
the Fermi velocity parallel to the c-axis.

       We use the general Kubo formula for the electronic thermal
conductivity of strong-coupling superconductors \cite{Ambegaokar} and
insert the Green's functions given in Eqs. (\ref{DefG}) and (\ref{DefF}).
For simplicity we consider the 2D-case where {\bf p} lies in the ab-plane
and the polar angle $\theta = \pi/2$. The integral over energy $\xi$ is
carried out by the method of residues (see Appendix C of 
Ref.~\onlinecite{Ambegaokar}) and yields the following result for 
$\kappa_{xx}$ \, $(x=\hat{\bf a})$ for ${\bf H}$ in the ab-plane:
\begin{eqnarray}
\kappa_{xx}
& = &
\frac{\pi\mbox{N}_{0}\mbox{v}^{2}}{8T^{2}}
\int_{0}^{\infty}d\omega\,\omega^{2}sech^{2}(\omega/2T)
\int_{0}^{2\pi}\frac{d\phi}{2\pi}\:\frac{2\cos^{2}\phi}{\mbox{Im}\xi_{0}}
\nonumber\\
& \times &
\frac{\left[1-\pi(\Delta\Lambda/\mbox{v})^{2}|f(\phi)|^{2}[\sin(\phi-\alpha)]^{-2}
|w(z_{0})|^{2}\right]}
{|1+2(\Delta\Lambda/\mbox{v})^{2}|f(\phi)|^{2}[\sin(\phi-\alpha)]^{-2}
\left[1+i\sqrt{\pi}z_{0}w(z_{0})\right]|^{2}}\, ,
\label{kapxx}
\end{eqnarray}
where
\begin{equation}
z_{0}=2(\omega+i\gamma)(\Lambda/\mbox{v})[\sin(\phi-\alpha)]^{-1} + 
i\sqrt{\pi}(\Delta\Lambda/\mbox{v})^{2}|f(\phi)|^{2}[\sin(\phi-\alpha)]^{-2}w(z_{0})\, ;
\label{z0a}
\end{equation}
\begin{equation}
z_{0}=(\omega+i\gamma+\xi_{0})(\Lambda/\mbox{v})[\sin(\phi-\alpha)]^{-1}\, .
\label{z0b}
\end{equation}
Here $\xi_{0}$ is the position of the pole of G in the upper half complex 
$\xi$-plane which is determined by the transcendental equation (\ref{z0a}).
For very low temperatures it is a good approximation to take the integrand
of the $\phi$-integral in Eq. (\ref{kapxx}) in the limit $\omega \rightarrow 0$.
Then the solution of Eq. (\ref{z0a}) for $z_{0}$ becomes purely imaginary:
$z_{0}=i x_{0}$. By using the boundaries for the w-function at imaginary
arguments one obtains the estimate \cite{BrandtThesis}
\begin{equation}
x_{0}=[2\gamma(\Lambda/\mbox{v})+\beta]/\sin(\phi-\alpha)\,;\quad
\mbox{Im}\xi_{0} = \gamma + \beta(\mbox{v}/\Lambda)\, ,
\label{x0}
\end{equation}
where
\begin{eqnarray}
\beta 
& = &
2(\Delta\Lambda/\mbox{v})^{2}|f(\phi)|^{2} \:
\{ \: [4(\gamma\Lambda/\mbox{v})^{2}
+4(\Delta\Lambda/\mbox{v})^{2}|f(\phi)|^{2}
+ \eta\sin^{2}(\phi-\alpha)]^{1/2}\nonumber\\
& &
\qquad\qquad\qquad\qquad+2(\gamma\Lambda/\mbox{v}) \:
\}^{-1}\:;\qquad (4/\pi\leq\eta\leq2)\, .
\label{beta}
\end{eqnarray}
Then the ratio of $\kappa_{xx}$ to the normal thermal 
conductivity $\kappa_{n}$ is approximately given by
\begin{eqnarray}
\frac{\kappa_{xx}}
{\kappa_{n}}
& = &
\int_0^{2\pi} \frac{d\phi}{2\pi} 2\cos^{2}(\phi)
\frac{\Gamma(\Lambda/\mbox{v})}{[(\gamma\Lambda/\mbox{v})+\beta]}
[1 - (\Delta\Lambda/\mbox{v})^{-2}|f(\phi)|^{-2}\beta^{2}]
\nonumber \\
& \times &
|1+2[\sin(\phi-\alpha)]^{-2}\{\:(\Delta\Lambda/\mbox{v})^{2}|f(\phi)|^{2}
-\beta[\beta+2(\gamma\Lambda/\mbox{v})]\:\}|^{-2}\, .
\label{kapratio}
\end{eqnarray}
\section{\quad Results for the Field Dependence of the thermal
conductivity for vertical and horizontal line nodes}

     We consider first the f-wave pairing state with vertical nodes at 
$\phi = \pm\pi/4$ and $\pm3\pi/4$ :  \cite{Hasegawa}
\begin{equation}
{\bf d}({\bf p}) = \Delta\,{\bf {\hat{z}}}\,(p_{x}+ip_{y})(p_{x}^{2}-p_{y}^{2})\, ;
\qquad|f(\phi)|^{2}=[\cos(2\phi)]^{2}\,.
\label{fwave}
\end{equation}
In Fig. 1 (a) and (b) we have plotted our results for 
$\kappa_{xx}/\kappa_{n}$ versus H/H$_{c2}$ for
$\Gamma/\Delta_{0}$=0.5\,(0.2) and $\beta_{A}=1.2$. Here we have used the
relationships (\ref{Defdellam}) connecting $(\Delta\Lambda/\mbox{v})^2$ and
$(\Lambda/\mbox{v})^2$ to the ratio H/H$_{c2}$. The upper solid curves in 
Fig. 1 refer to the case ${\bf H}\parallel\bf {\hat{c}}$ which, according to 
Eq. (\ref{sin}), is obtained from Eq. (\ref{kapratio}) by replacing 
$\sin(\phi-\alpha)$ by $\sin\theta$ with $\theta=\pi/2$. The lower dashed 
curves show the case ${\bf H}\parallel\bf {\hat{a}}$, i.e., $\alpha=0$. The 
solid curve in Fig. 1(a) increases almost linearly with H and agrees roughly 
with the data for perpendicular field shown in Fig. 1(a) of 
Ref.~\onlinecite{Izawa} at the lowest
temperature. The solid curve in Fig. 1(b) for $\Gamma/\Delta_{0}=0.2$
disagrees with the data because of its upward curvature. The dashed curves
both exhibit an upward curvature near H$_{c2}$ which is however stronger in 
Fig. 1(b). Fig. 1(b) is in fact in better agreement with the data for parallel field 
shown in Fig. 1(b) of Ref.~\onlinecite{Izawa} due to the rapid increase of the 
thermal conductivity as H approaches H$_{c2}$.

     We next consider the angular variation of the thermal conductivity in a
parallel field. In Fig. 2(a) we show an example of 
$\kappa_{xx}/\kappa_{n}$ vs the angle $\alpha$ between {\bf H} and 
the a-axis for $(\Delta\Lambda/$v)=0.2 and $\Gamma/\Delta_{0}=0.1$. The
angular variation is seen to exhibit two-fold symmetry proportional to
$\cos(2\alpha)$ with the minimum at $\alpha=0$ where {\bf H} is parallel to
the heat current and the maximum at $\alpha=\pi/2$. This behavior can be
explained as an effect of the density of states which has a minimum for 
quasiparticles travelling parallel to the field and a maximum for quasiparticles
moving perpendicular to the field.\cite{Brandt} In fact, the expression
$|\cdots|^2$ in the denominators of equations (\ref{kapxx}) and (\ref{kapratio})
corresponds to $|N(\phi,\alpha)|^2$. In Fig. 3(a) we show the relative amplitude
$\Delta\kappa = 2[\kappa(\alpha=\pi/2) - \kappa(\alpha=0)]/
[\kappa(\pi/2)+\kappa(0)]$ for $\Gamma/\Delta_{0}=0.1$ (solid curve) and 
0.5 (dashed curve) vs H/H$_{c2}$. One sees that this amplitude becomes
small for H/H$_{c2}\rightarrow 1$ and that it decreases with increasing
impurity scattering rate $\Gamma/\Delta_{0}$. However, even for
$\Gamma/\Delta_{0}=0.5$, the amplitudes are still larger percentages of 
$\kappa$ than those shown in Fig. 2(b) of Ref.~\onlinecite{Izawa}.

    We have also considered the case where the heat current is in the direction
$\phi_{\kappa}=\pi/4$ corresponding to the data for ${\bf q}\parallel$ [110]
shown in Fig. 2(a) of Ref.~\onlinecite{Izawa}. Then the directional factor
$2\cos^{2}(\phi)$ in Eq. (\ref{kapratio}) has to be replaced by
$2\cos^{2}(\phi-\pi/4)=1+\sin(2\phi)$. As an example we show in Fig. 2(a) 
$\kappa_{s}/\kappa_{n}$ vs $\alpha$ for $\Delta\Lambda/$v=0.2 and
$\Gamma/\Delta_{0}=0.1$. We see again the two-fold symmetry with the 
minimum occuring at $\alpha=\pi/4$, i.e., for {\bf H} parallel to the heat current
{\bf q}, and the maximum at $\alpha=3\pi/4$. This behavior is basically in
agreement with the data in Fig. 2(a) of Ref.~\onlinecite{Izawa}. However, at
higher fields for $\Delta\Lambda/$v=0.1 (see Fig. 2(b)), a small local maximum
occurs at  $\alpha=\pi/4$ while a large maximum remains at $\alpha=3\pi/4$.
The occurence of a local maximum and two neighboring minima is plausible
since the quasiparticles are now travelling mainly along the node of the gap
where the effect of the field on the density of states vanishes. This behavior
seems to differ from the data in Fig. 2(a) of Ref.~\onlinecite{Izawa} at higher
fields where a small component of 4-fold symmetry with minimum for 
${\bf H}\parallel{\bf q}$ has been extracted 
(see Fig. 3 of Ref.~\onlinecite{Izawa}). The order of magnitude of the relative
amplitudes of the calculated oscillations for the heat current along the node
of the gap is about the same as that shown in Fig. 3(a) where the heat current
is directed along the antinode of the gap.

     In Fig. 2(b) we have plotted our results for 
$\kappa_{s}/\kappa_{n}$ vs $\alpha$ for the case where the factor
$2\cos^{2}(\phi)$ in Eq. (\ref{kapratio}) has been replaced with 1. The 
resulting curve is seen to have 4-fold symmetry in $\alpha$ with minima
at $\alpha=0$ and $\alpha=\pi/2$ and maxima at $\alpha=\pi/4$ and
$\alpha=3\pi/4$. This form is similar to the 4-fold symmetry of $\kappa$
observed experimentally in YBCO. \cite{Aubin} The explanation may be 
the following: According to the theory of thermal  conductivity,
\cite{Ambegaokar} the conductivity $\kappa_{\mbox{ii}}$ along axis i 
contains the factor v$_{\mbox{i}}^{2}$ where v$_{\mbox{i}}$ is the 
component of the Fermi velocity along the i-axis. For a circular Fermi 
line this yields a factor proportional to $\cos^{2}(\phi-\phi_{i})$. However, 
for YBCO the Fermi line is roughy a square and thus the component 
v$_{\mbox{b}}$ along the b-axis is approximately constant along each side 
of the square. It is interesting that the data for Sr$_2$RuO$_4$ yield a 
small component of 4-fold symmetry which is similar to that shown in 
Fig. 2(b). This has been attributed however to the 4-fold anisotropy of 
H$_{c2}$.\cite{Izawa} For lower fields we find an interchange of the 
maxima and minima of the 4-fold variation which is in agreement with 
the results in Ref.~\onlinecite{Dahm} (see Fig. 2(a)).

     We turn now to the f-wave pairing state with vertical nodes where the term
$(p_{x}^{2}-p_{y}^{2})$ in Eq. (\ref{fwave}) is replaced by $p_{x}p_{y}$
and thus the squared amplitude $[\cos(2\phi)]^{2}$ in Eq. (\ref{fwave}) 
is replaced by $[\sin(2\phi)]^{2}$.  \cite{Hasegawa} The resulting thermal
conductivity is simply obtained from Eq. (\ref{kapratio}) for state (\ref{fwave})
by making a variable transformation $\phi = \phi^{\prime} - \pi/4$ in the 
$\phi$-integral which yields the new field direction angle 
$\alpha^{\prime} = \alpha + \pi/4$ and the heat current direction 
$\phi_{\kappa}=\pi/4$. Thus we obtain for the new state the same function of
angle $\alpha^{\prime}$ as the function of $\alpha$ in Fig. 2(a) (solid curve) 
shifted by 
$\pi/4$. Vice versa, for the heat current along the node of the old state the
variable transformation $\phi = \phi^{\prime} + \pi/4$ yields for the new state the
same functions of the field angle $\alpha^{\prime} = \alpha - \pi/4$ as in 
Fig. 2(a) (dash-dot curve) shifted by $-\pi/4$ where the heat current is now along 
$\phi_{\kappa}=0$. The latter form of $\kappa(\alpha)$ agrees with that which
has been obtained by the Doppler shift method for low fields in 
Ref.~\onlinecite{Dahm}.

     The authors of Ref.~\onlinecite{Izawa} suggested that a superconducting 
state with horizontal line nodes is a better candidate for explaining their
thermal conductivity data. We therefore now consider the state 
 \cite{Hasegawa}
\begin{equation}
{\bf d}({\bf p}) = \Delta{\bf {\hat{z}}}(p_{x}+ip_{y})[\cos(cp_{z}) + a_{0}]\, ;
\qquad|f|^{2}=[\cos(cp_{z}) + a_{0}]^{2}\,.
\label{fwave2}
\end{equation}
where $a_{0}\leq 1$. Now the expression for 
$\kappa_{xx}/\kappa_{n}$ in
Eq. (\ref{kapratio}) has to be integrated over $p_{z}$. This leads to an 
additional integral over the new variable $\theta = cp_{z}$ from $-\pi$
to $+\pi$ of the expression in Eq. (\ref{kapratio}) with the squared gap
amplitude replaced with $|f|^{2}=(\cos(\theta) + a_{0})^{2}$. For 
${\bf H}\parallel\bf {\hat{c}}$ the terms $\sin(\phi - \alpha)$ in equations
(\ref{beta}) and (\ref{kapratio}) have to be replaced by 1 which yields for
$a_{0} = 0$ the same result for $\kappa_{xx}/\kappa_{n}$ as that
for state (\ref{fwave}). For {\bf H} lying in the ab-plane one has to carry 
out the double integral over $\theta$ and $\phi$. In Fig. 4(a) we show 
$\kappa_{xx}/\kappa_{n}$ vs H/H$_{c2}$ for ${\bf H}\parallel\bf {\hat{a}}$
and parameter values $\Gamma/\Delta_{0}$=0.5 (0.1) and $\beta_{A}=1.2$.
The results are seen to be very similar to the dashed curves in Figs. 1(a) and
1(b) for the state (\ref{fwave}). The modulation of $\kappa_{xx}/\kappa_{n}$
with rotation angle $\alpha$ has two-fold symmetry with a minimum at 
$\alpha=0$ and a maximum at $\alpha=\pi/2$ as can be seen in Fig. 4(b). 
The plots of $\kappa_{xx}/\kappa_{n}$ vs $\alpha$ are very similar to those
for state (\ref{fwave}) (see Fig. 2(a), solid curve). In Fig. 3(b) we show our 
results for the relative amplitudes of the modulations for impurity scattering 
$\Gamma/\Delta_{0}$=0.5 and 0.1. These are nearly the same as those for
state (\ref{fwave}) shown in Fig. 3(a). The main difference between the states
with horizontal and vertical line nodes is that, for the former states, the
two-fold modulation of $\kappa$ with rotation angle $\alpha$ of the in-plane
field is always the same function relative to the direction of the heat current
while, for the latter states, the modulation depends somewhat on the direction
of the heat current relative to the vertical nodes (see Figs. 2(a) and 2(b)). This
can be seen from Eq. (\ref{kapratio}) with the variable transformation
$\phi = \phi^{\prime} - \phi_{\kappa}$ for the heat current in the direction 
$\phi_{\kappa}$ which leads to the same function of the new rotation angle 
$\alpha^{\prime} = \alpha + \phi_{\kappa}$ since $|f|^{2} = \cos^{2}\theta$ does 
not depend on $\phi$. Thus we see that our results for states with horizontal 
nodes, as well as for states with vertical nodes, agree roughly with the
observed thermal conductivity data. This is because the minimum of $\kappa$
in Figs. 2(a) and 2(b) of Ref.~\onlinecite{Izawa} for heat current 
${\bf q}\parallel$ [110] and ${\bf q}\parallel$ [100] occurs in both cases for
${\bf H}\parallel{\bf q}$.

     We have also calculated the thermal conductivity for the following f-wave
pairing state with horizontal line nodes which has been proposed for UPt$_3$:  
\cite{WonMaki}
\begin{equation}
{\bf d}({\bf p}) = (3\sqrt{3}/2)\Delta\,{\bf {\hat{z}}}\,p_{z}(p_{x}+ip_{y})^{2}\, ;
\qquad|f|^{2}=(27/4)\cos^{2}\theta(1 - \cos^{2}\theta)^{2}\,.
\label{fwave3}
\end{equation}
The corresponding expression for $\kappa_{bb}/\kappa_{n}$ is obtained from
Eq. (\ref{kapratio}) by replacing the directional terms $\sin(\phi-\alpha)$ due to
the field by the expressions in Eq. (\ref{sin}) with v$^{\prime}$=v. The directional
terms $|f|^2$ due to the squared gap amplitude are given now by 
Eq. (\ref{fwave3}). Additionally, the heat flow directional term $2\cos^{2}\phi$
in Eq. (\ref{kapratio}) is replaced by $3\sin^{2}\theta\sin^{2}\phi$ for heat flow
along the b-axis. The integrations over the polar angle $\theta$ and the 
azimuthal angle $\phi$ yield the variation of $\kappa_{bb}$ vs the rotation angle
$\alpha$ of {\bf H} in the ab-plane shown in Fig. 4(b). One see that the 
modulation of $\kappa_{bb}(\alpha)$ has two-fold symmetry in $\alpha$ with a
minimum at ${\bf H}\parallel\bf {\hat{b}}\parallel{\bf q}$ ({\bf q} is the direction
of the heat flow) and a maximum at ${\bf H}\parallel\bf {\hat{a}}$. These results
are contrary to the two-fold variation of $\kappa$ observed in UPt$_3$ which
shows a maximum for the field parallel to the heat flow. \cite{Suderow} In 
Fig. 4(a) we show the dependence of $\kappa_{bb}$ on H/H$_{c2}$ for
$\alpha=\pi/2$, i.e., ${\bf H}\parallel\bf {q}\parallel{\bf {\hat{b}}}$. One sees that
$\kappa_{bb}$ exhibits a strong upward curvature for small impurity scattering
rate $\Gamma/\Delta_{0} = 0.1$ while it becomes approximately linear in H 
for the larger scattering rate $\Gamma/\Delta_{0} = 0.5$. According to the
measurements of $\kappa$ for UPt$_{3}$, the thermal conductivity $\kappa_{bb}$
for the configuration ${\bf H}\parallel\bf {q}\parallel{\bf {\hat{b}}}$ increases 
almost linearly with H while, for the configuration 
${\bf H}\parallel\bf {q}\parallel{\bf {\hat{c}}}$, $\kappa_{cc}$ exhibits a steep
increase as H tends to H$_{c2}$. \cite{Suderow} Therefore we have also
calculated  $\kappa_{cc}$ for ${\bf H}\parallel\bf {q}\parallel{\bf {\hat{c}}}$ and
find surprisingly nearly the same field dependence as for $\kappa_{bb}$, i.e.,
in both cases $\kappa$ is linear in H or increases steeply at H$_{c2}$. Thus 
we see that both our results for the field dependence of $\kappa$ for 
${\bf H}\parallel\bf {\hat{b}}$ and ${\bf H}\parallel\bf {\hat{c}}$, and the
variation of $\kappa(\alpha)$ with the in-plane field rotation angle $\alpha$ are
in disagreement with the measurements on UPt$_{3}$. \cite{Suderow}

     Finally we discuss the $\omega$-dependence of the integrand of
Eq. (\ref{kapxx})  which determines the temperature dependence of the thermal 
conductivity.  It is now necessary to solve Eq. (\ref{z0a}) for $z_{0}$ as a
function of $\omega$ and to integrate the resulting expression in
Eq. (\ref{kapxx}) over $\phi$. In Fig. 5(a) we have plotted the result 
(without the factor $\omega^{2}sech^{2}(\omega/2T)\,$) vs 
$\Omega = \omega/\Delta$ for the state (\ref{fwave}) at field directions
 $\alpha=0$ and $\alpha=\pi/2$ for parameter values $\Delta\Lambda/v = 0.6$ 
and $\Gamma/\Delta_{0}=0.2$. The results for $\omega=0$ agree with those 
obtained previously from Eq. (\ref{kapratio}). Most interesting is the result that 
the curves for $\alpha=0$ and $\pi/2$ in Fig.5(a) cross each other at about 
$\Omega\simeq 0.7$ which means that for higher frequencies the former 
minimum of the integrand of $\kappa_{xx}$ becomes a maximum and 
vice versa. This leads to a reduction of the field modulation at finite 
temperatures due to the factor $\omega^{2}sech^{2}(\omega/2T)$ in the integrand
of Eq. (8). In Fig. 5(b) we show our result for the temperature dependence of the
relative amplitude $\Delta\kappa$ which is obtained by carrying out the
$\omega$-integrals in Eq. (\ref{kapxx}) with the integrands shown in Fig. 5(a). 
One recognizes that $\Delta\kappa$ decreases rapidly with increasing 
T/T$_{C}$ and becomes small of the order of a few percent at the experimental 
values of T/T$_{C}$. \cite{Izawa} Furthermore note that $\Delta\kappa$ even 
becomes negative in the temperature range between T/T$_{C}\simeq0.25$ and
0.75 which means that the minimum and maximum are interchanged.

     We have also calculated the frequency dependence of the integrand of
$\kappa_{xx}$ in Eq. (\ref{kapxx}) with the factor $2\cos^2\phi$ replaced by 1 
which yields the 4-fold variation of $\kappa(\alpha)$ discussed above. For 
lower fields, or larger $\Delta\Lambda/\mbox{v}$, the maximum and minimum
of $\kappa(\alpha)$ occur, for $\omega=0$, at field directions $\alpha=0$ and
$\alpha=\pi/4$, respectively. In Fig. 6 we show the curves $\kappa(\Omega)$
for $\alpha=0$ and $\alpha=\pi/4$ for the same parameter values as in 
Fig. 5(a). Note that the curves cross each other for increasing frequency at a 
much lower frequency ($\Omega\simeq0.3$) than in Fig. 5(a). This means
that the maxima and minima of $\kappa(\alpha)$ are interchanged at a much
lower temperature than that in Fig. 5(b). This is in agreement with the
measurements on YBCO. \cite{Aubin}

\section{\quad Conclusions}

     In summary, we have calculated the in-plane thermal conductivity 
$\kappa$ near H$_{c2}$ for different spin-triplet pairing states with line 
nodes in the superconducting gap which are directed 
perpendicular (vertical) and parallel (horizontal)to the ab-plane. For fields 
along the c-axis, $\kappa$ increases approximately linearly with 
increasing field, while, for fields parallel to the ab-plane, $\kappa$ exhibits 
an upward curvature as H approaches H$_{c2}$. For rotating in-plane 
magnetic field we obtain variations $\kappa(\alpha)$ of
two-fold symmetry in the rotation angle $\alpha$ where the minimum and
maximum occur for fields parallel and perpendicular to the heat flow. The
relative amplitude $\Delta\kappa$ of the modulation decreases with
increasing field, impurity scattering, and temperature. At higher temperatures
$\Delta\kappa$ even changes sign indicating that the minimum and 
maximum of  $\kappa(\alpha)$ are interchanged. Since all these results are 
nearly the same for the spin-triplet pairing models with  line nodes vertical 
or horizontal to the RuO$_{2}$ planes, we cannot decide which one of the
proposed models is more compatible with the recent data for 
Sr$_2$RuO$_4$. \cite{Izawa,Tanatar} The smallness of the observed 
variations of the in-plane $\kappa$ for rotating in-plane field may well be 
accounted for by finite temperature effects. The only marked difference 
between our results for these two types of states is that for horizontal nodes 
the variation $\kappa(\alpha)$ always has the same form relative to the 
direction of the heat current while, for vertical nodes $\kappa(\alpha)$ is 
somewhat different for the heat current in the direction of an antinode or a 
node. For the heat current in the direction of a vertical node $\kappa(\alpha)$
shows a 2-fold symmetric modulation for lower fields (see the dash-dot
curve in Fig. 2(a)). A similar form has been obtained in 
Ref.~\onlinecite{Dahm} for the state with squared amplitude
$\sin^2(2\phi)$. We have shown in Section III that, for this state, the
dash-dot curve in Fig. 2(a) is shifted by $-\pi/4$. For higher fields this
modulation $\kappa(\alpha)$ develops some structure for field directions
near $\alpha=\pi/4$ corresponding to the direction of the node (see the
solid curve in Fig. 2(b)). Therefore we suggest more accurate 
measurements of $\kappa(\alpha)$ for different directions of the heat 
flow which could distinguish between the two types of pairing states. 
It would also be interesting to measure $\kappa(\alpha)$ at higher 
temperatures to see whether the predicted interchange of minima and 
maxima actually takes place. 

     No significant anisotropy in the inter-plane $\kappa$ of 
Sr$_2$RuO$_4$ was observed for rotating in-plane field, perhaps 
indicating the existence of horizontal line nodes. \cite{Tanatar} In fact, 
for the states (15) and (16) with horizontal line nodes, $\kappa_{cc}$ 
along the c-axis actually does not depend on the in-plane rotation 
angle $\alpha$.

     The f-wave pairing state with horizontal line nodes, which has been
discussed for Sr$_2$RuO$_4\,$, has also been proposed for UPt$_{3}$.
 \cite{WonMaki} We have calculated $\kappa_{s}$ with the help
of the three-dimensional {\bf p}-integral and find that, for relative
orientations  ${\bf H}\parallel{\bf \mbox{q}}\parallel\bf {\hat{\mbox{b}}}$
as well as ${\bf H}\parallel{\bf \mbox{q}}\parallel\bf {\hat{\mbox{c}}}$,
the field dependence of $\kappa_{s}$ is nearly the same: it
exhibits a steep upward curvature near H$_{c2}$ for low impurity
scattering rate $\Gamma/\Delta_0$ while it becomes more linear for
larger values of $\Gamma/\Delta_0$. These results disagree with
measurements in  UPt$_{3}$ where $\kappa$ is found to be almost
linear in H for the configuration  
${\bf H}\parallel{\bf \mbox{q}}\parallel\bf {\hat{\mbox{b}}}$ while it 
increases steeply near H$_{c2}$ for
${\bf H}\parallel{\bf \mbox{q}}\parallel\bf {\hat{\mbox{c}}}$. 
 \cite{Suderow} For rotation of {\bf H} in the ab-plane we obtain a
modulation $\kappa(\alpha)$ of two-fold symmetry in the angle
$\alpha$ with a minimum for ${\bf H}\parallel\bf {\hat{\mbox{b}}}$
and a maximum for ${\bf H}\parallel\bf {\hat{\mbox{a}}}$. This
result is in agreement with that obtained by the Doppler shift
method for low fields. \cite{WonMaki} However, these results are
contrary to the observed two-fold variation of $\kappa(\alpha)$ in
UPt$_{3}$ which has a maximum for the field parallel to the heat
flow. \cite{Suderow} Since these measurements were carried out at 
higher temperatures it may be that the interchange of maximum 
and minimum is due to the finite temperature effect. However, the
disagreement between theory and experiment discussed above
casts some doubt on the validity of this f-wave pairing model for
UPt$_3$.

    Our calculation of $\kappa$ for the f-wave pairing state with
vertical line nodes (squared gap amplitude $\cos^2(2\phi)$) also
applies to the d$_{x^2-y^2}\:$ -wave pairing state which is widely 
believed to be realized in the high-T$_{C}$ cuprates. It is 
interesting that we obtain agreement with the observed four-fold 
symmetric variation $\kappa(\alpha)$ in YBCO  \cite{Aubin} if we 
neglect the directional
factor arising from the squared component of the Fermi velocity in
the direction of the heat flow. This approximation might be valid for
the nearly square Fermi surface of YCBO because the Fermi velocity
is almost constant on each side of the square. Our result that, for 
lower  fields, the minima and maxima in $\kappa(\alpha)$ are
interchanged is a problem that also occurs in the Doppler shift
method at lower fields. \cite{Dahm} Here again, the finite frequency
behavior, as seen in Fig. 6, and consequently the finite temperature
effect calculated at the end of Section III, are responsible for the
interchange of minima and maxima in the 4-fold variation 
$\kappa(\alpha)$ of the thermal conductivity which leads to
agreement with the measurements on YBCO. \cite{Aubin} In general,
this interchange of maxima and minima is quite sensitive to the
strength of the magnetic field and to the temperature.

     Resolution of the remaining descrepancies between experiment
and theory will require more measurements of the thermal conductivity,
in particular, for different directions of the heat flow. It should be
pointed out that we have made some approximations in our theory of
thermal conductivity \cite{Ambegaokar} near H$_{c2}$. \cite{Brandt}
First, we have neglected the small contributions of higher Landau
levels N to the Abrikosov order parameter which occur for
spin-triplet pairing. \cite{Scharnberg} Second, we have neglected higher
order Fourier coefficients of the Green's functions with respect to
wave vectors {\bf k} of the reciprocal Abrikosov vortex lattice
 \cite{Brandt} which become more and more important as the applied 
field is decreased. At fields near H$_{c1}$ the complementary method 
of the Doppler effect for a single vortex becomes more appropriate, 
however, there the scattering of quasiparticles by a vortex core is 
neglected whereas it is taken into account in an average way by our 
method.

\acknowledgements

We would like to thank T. Dahm for helpful advice and fruitful
discussions.	
\newpage
\newpage
\vspace{0.5in}
  
FIGURE\ CAPTIONS 

1. Thermal conductivity $\kappa_{s}/\kappa_{n}$ vs applied
field H/H$_{c2}$ at T=0 for vertical line nodes (state \ref{fwave}) for {\bf H}
parallel to the c-axis (solid curves) and {\bf H} parallel to the 
a-axis (dashed curves) for different impurity 
scattering parameters: (a) $\Gamma/\Delta_0$=0.5; (b) $\Gamma/\Delta_0$=0.2.
The Abrikosov parameter is taken as $\beta_A$=1.2.

2.  $\kappa_{\mbox{s}}/\kappa_{\mbox{n}}$ vs 
$\alpha = \angle\bf {\mbox{(H,a)}}$ for
in-plane field rotation at T=0 for state (\ref{fwave}) with vertical nodes and 
impurity scattering rate $\Gamma/\Delta_0$=0.1. (a) Heat current {\bf q} in the 
direction of a vertical antinode (solid curve), {\bf q} in the direction of a vertical 
node (dash-dot curve), and no directional factor for {\bf q} (dashed curve), all 
for gap parameter $\Delta\Lambda$/v=0.2; (b) {\bf q} in the direction of a node
(solid curve), and no directional factor for {\bf q} (dashed curve), both for
$\Delta\Lambda$/v=0.1.

3. Relative amplitude $\Delta\kappa$ of the variation of $\kappa(\alpha)$ for
rotating in-plane fields vs H/H$_{c2}$ at T=0. (a) State (\ref{fwave}) with vertical 
nodes and $\Gamma/\Delta_0$=0.1 (solid curve) and $\Gamma/\Delta_0$=0.5 
(dashed curve). (b) State (\ref{fwave2}) with horizontal line nodes and
$\Gamma/\Delta_0$=0.1 (solid curve) and $\Gamma/\Delta_0$=0.5 (dashed 
curve).

4. Magnetic field dependence of $\kappa_{\mbox{s}}$ for horizontal line 
nodes at T=0.
(a) $\kappa_{\mbox{s}}/\kappa_{\mbox{n}}$ vs H/H$_{c2}$ for state 
(\ref{fwave2}) with ${\bf H}\parallel{\bf \mbox{q}}\parallel\bf {\hat{\mbox{a}}}$ 
(dashed curves), and for state (\ref{fwave3}) with 
${\bf H}\parallel{\bf \mbox{q}}\parallel\bf {\hat{\mbox{b}}}$ (solid curves), for
$\delta=\Gamma/\Delta_0$=0.1 and 0.5; 
(b) $\kappa_{s}/\kappa_{n}$ vs $\alpha = \angle\bf {\mbox{(H,a)}}$ for
state (\ref{fwave2}) with {\bf q} $\parallel\bf {\hat{\mbox{a}}}$ (solid curve), and
for state (\ref{fwave3}) with {\bf q} $\parallel\bf {\hat{\mbox{b}}}$ (dashed curve),
for parameter values $\Delta\Lambda$/v=0.2 and $\Gamma/\Delta_0$=0.1.

5. (a) Integrand in Eq. (\ref{kapxx}) for $\kappa_{s}/\kappa_{n}$ 
(without the factor $\omega^{2}sech^{2}(\omega/2T)$) vs reduced frequency
$\Omega=\omega/\Delta$ for state (\ref{fwave}) and gap parameter 
$\Delta\Lambda$/v=0.6, for field direction $\alpha=0$ (solid curve) and
$\alpha=\pi/2$ (dashed curve). The impurity scattering rate is
$\Gamma/\Delta_0$=0.2.
(b) Relative amplitude $\Delta\kappa$ of the variation of $\kappa(\alpha)$
for state (\ref{fwave}) vs T/T$_C$, for gap parameter 
$\Delta\Lambda$/v=0.6 and impurity scattering rate $\Gamma/\Delta_0$=0.2
(solid curve) and $\Gamma/\Delta_0$=0.5 (dashed curve).

6. Integrand in Eq. (\ref{kapxx}) for $\kappa_{s}/\kappa_{n}$ 
(without the factor $\omega^{2}sech^{2}(\omega/2T)$) and no directional
factor for heat flow (corresponding to YCBO) vs reduced frequency
$\Omega=\omega/\Delta$ for state (\ref{fwave}) and gap parameter 
$\Delta\Lambda$/v=0.6, for field direction $\alpha=0$ (solid curve) and
$\alpha=\pi/4$ (dashed curve). The impurity scattering rate is
$\Gamma/\Delta_0$=0.2.

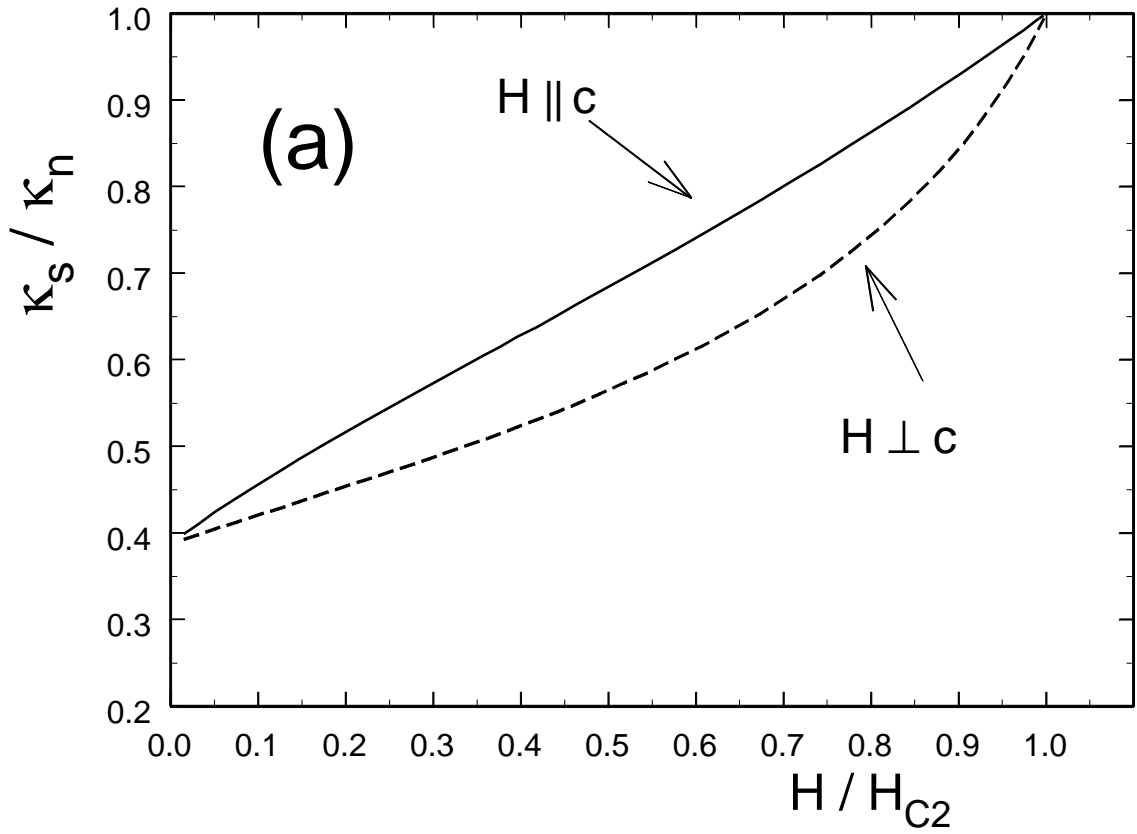
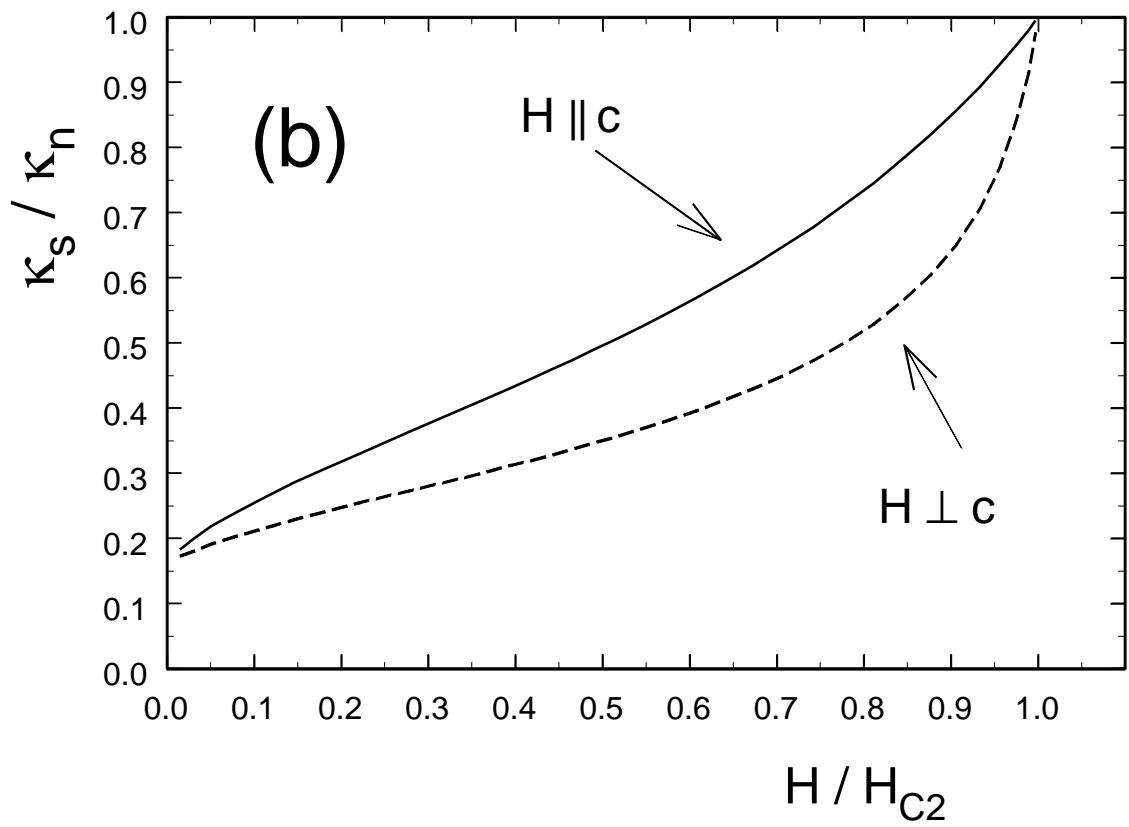
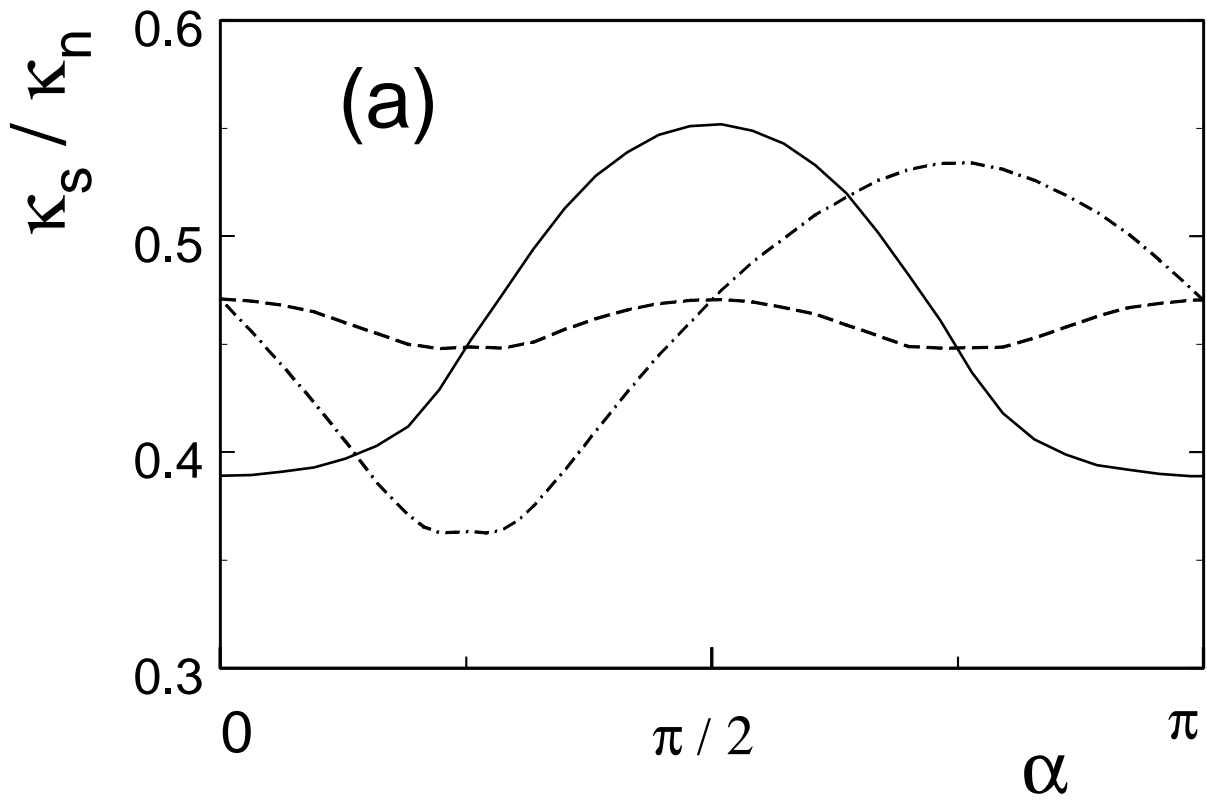
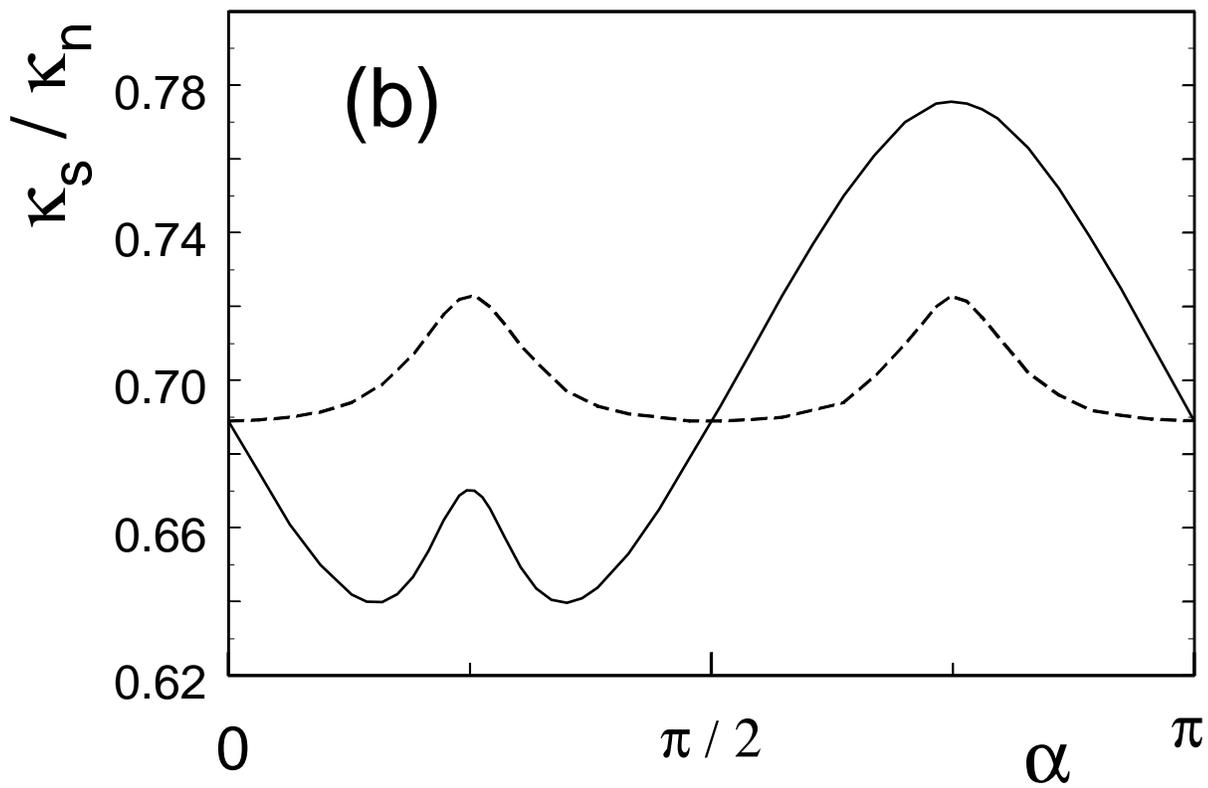
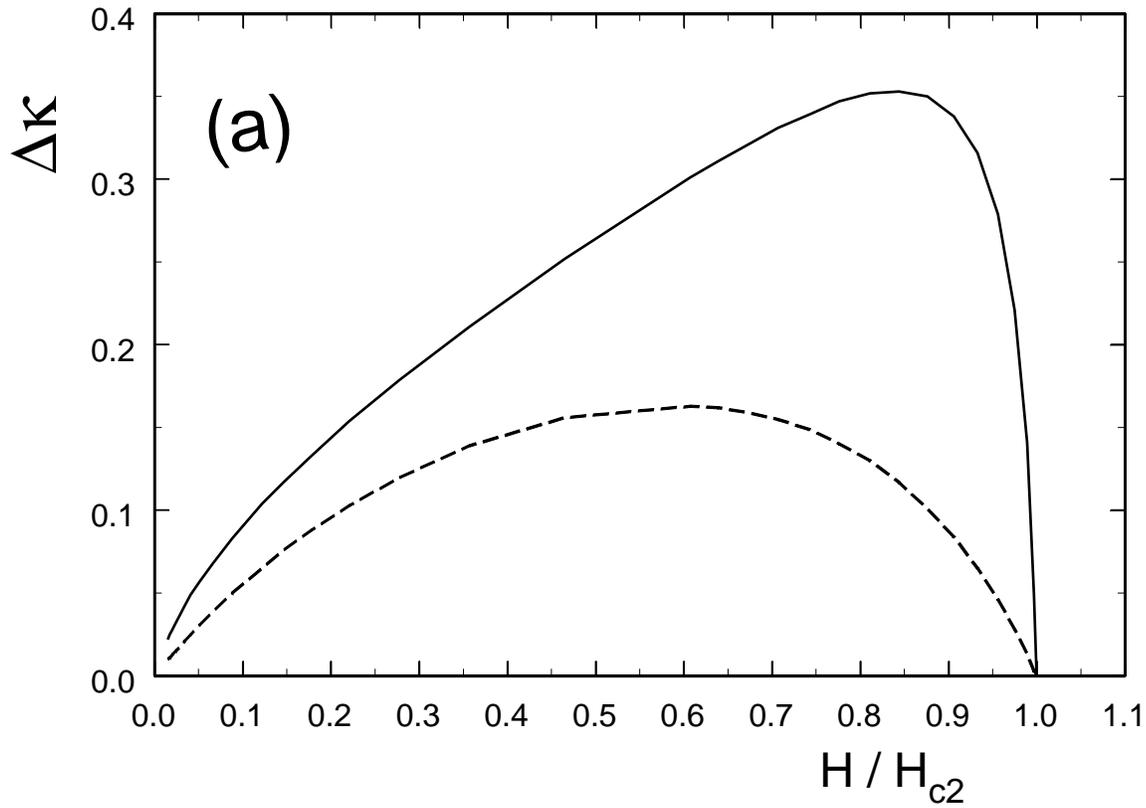
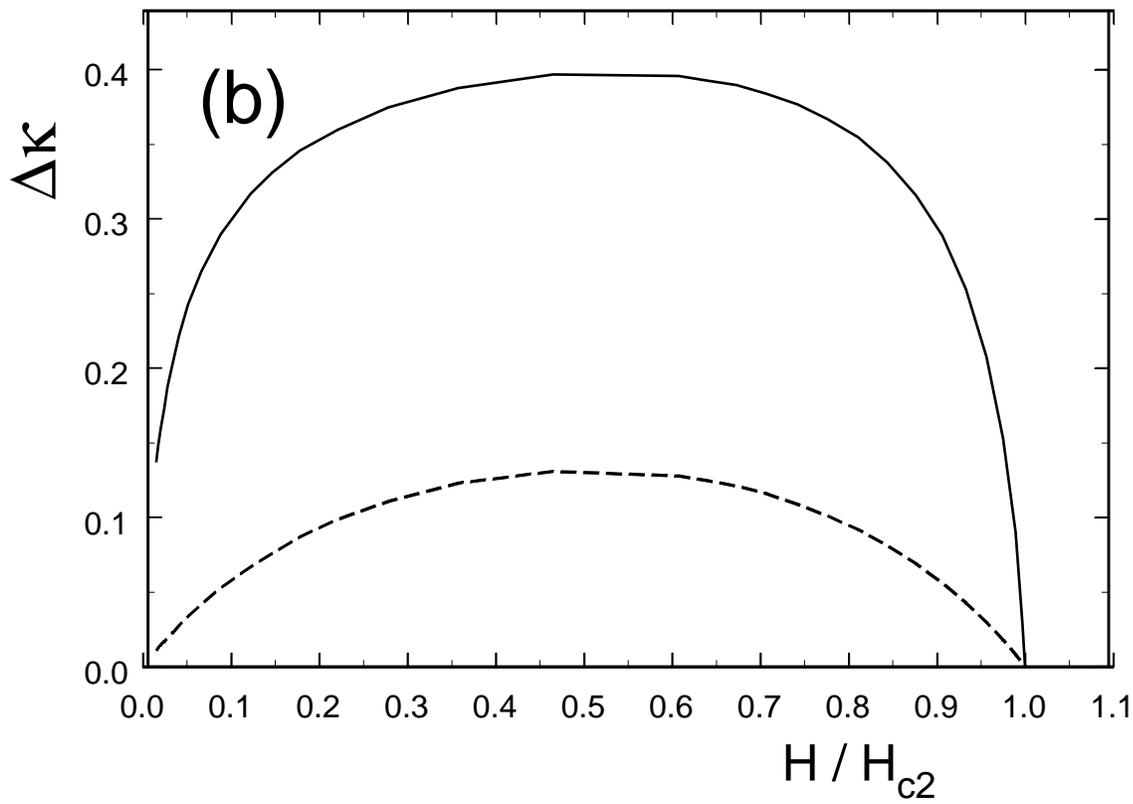
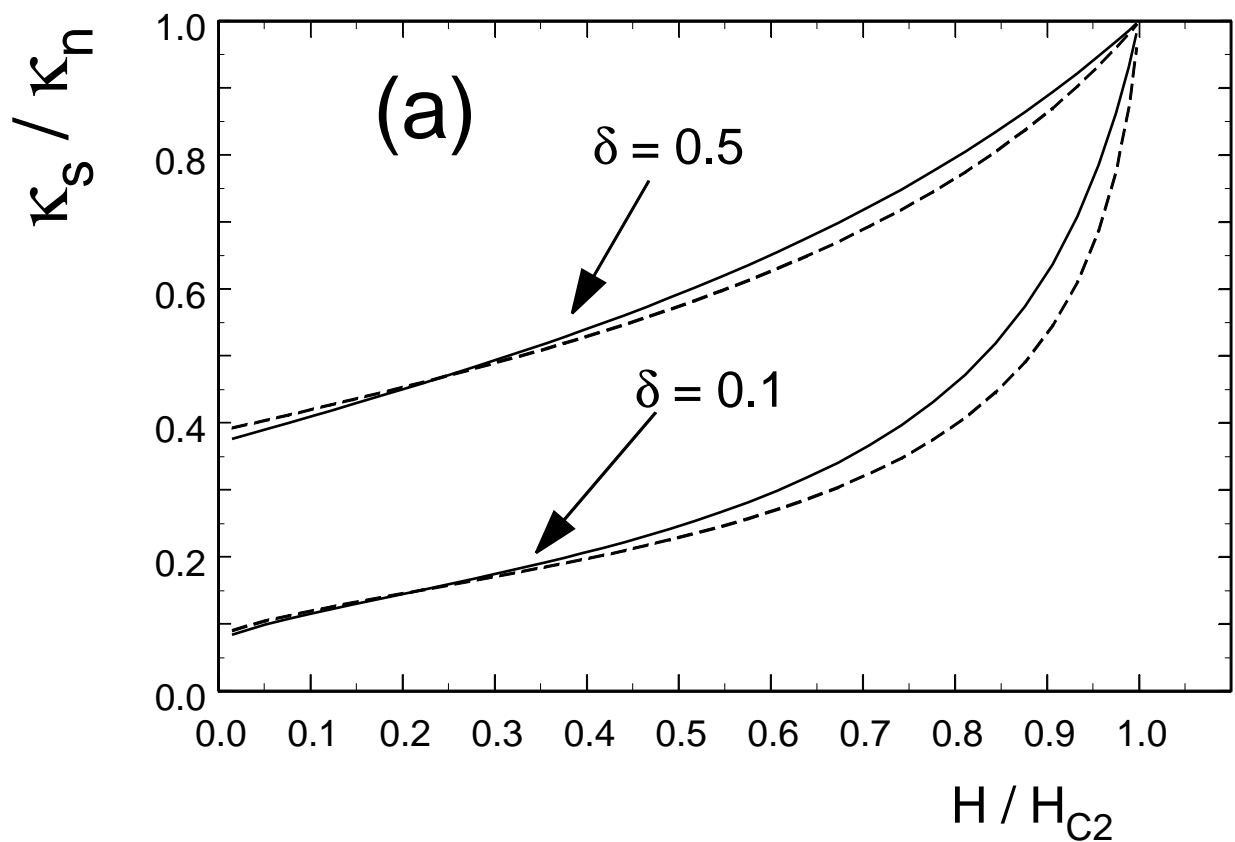
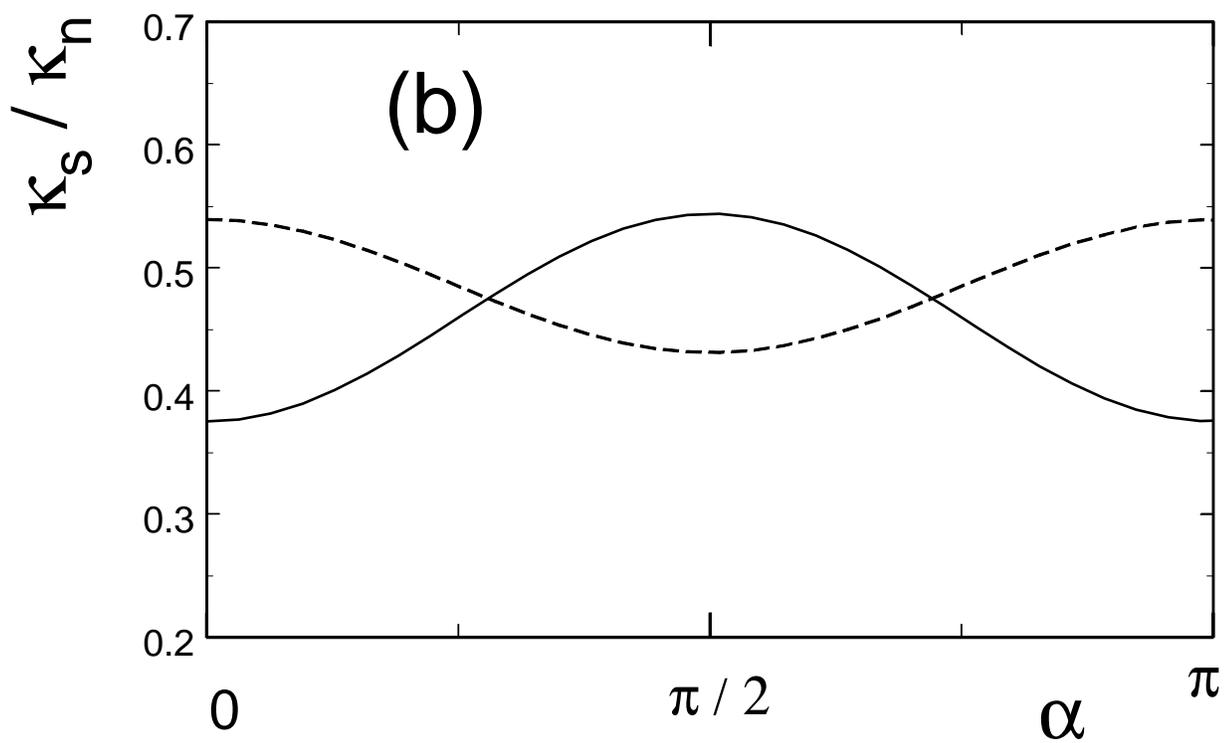
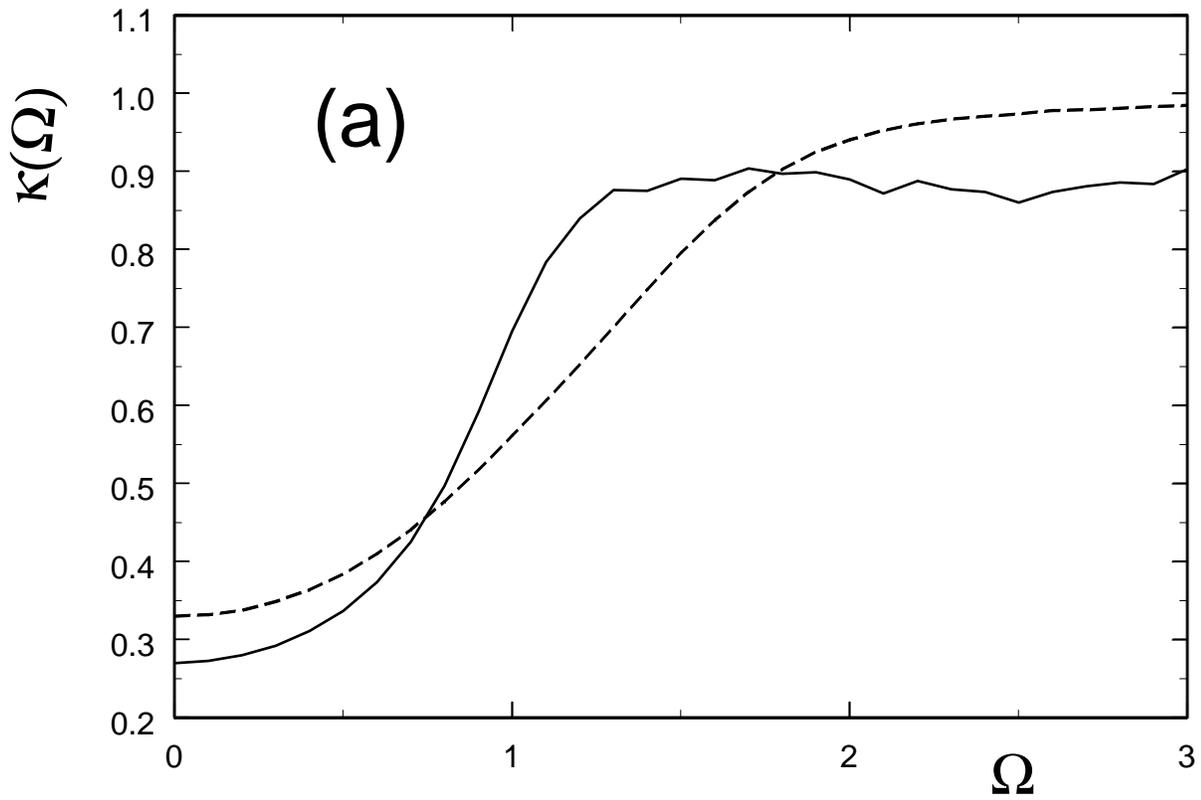
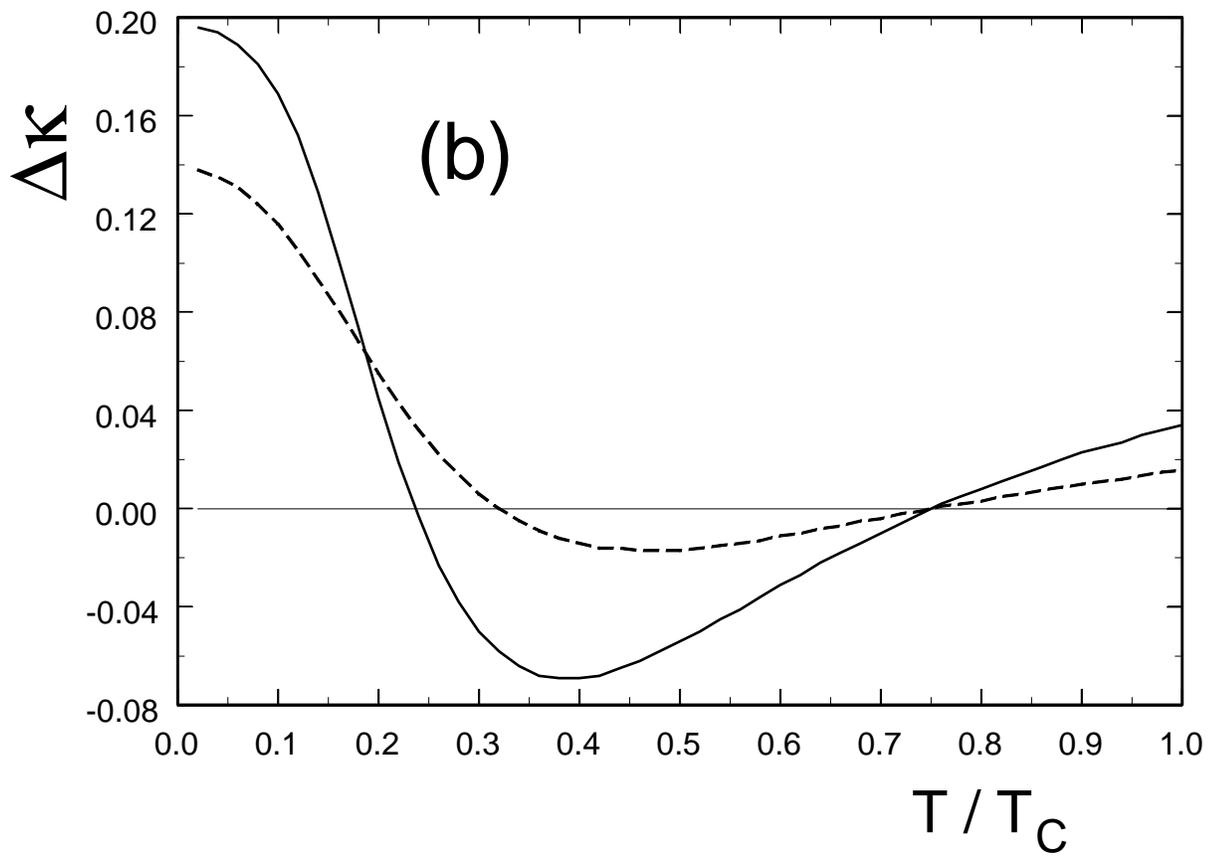
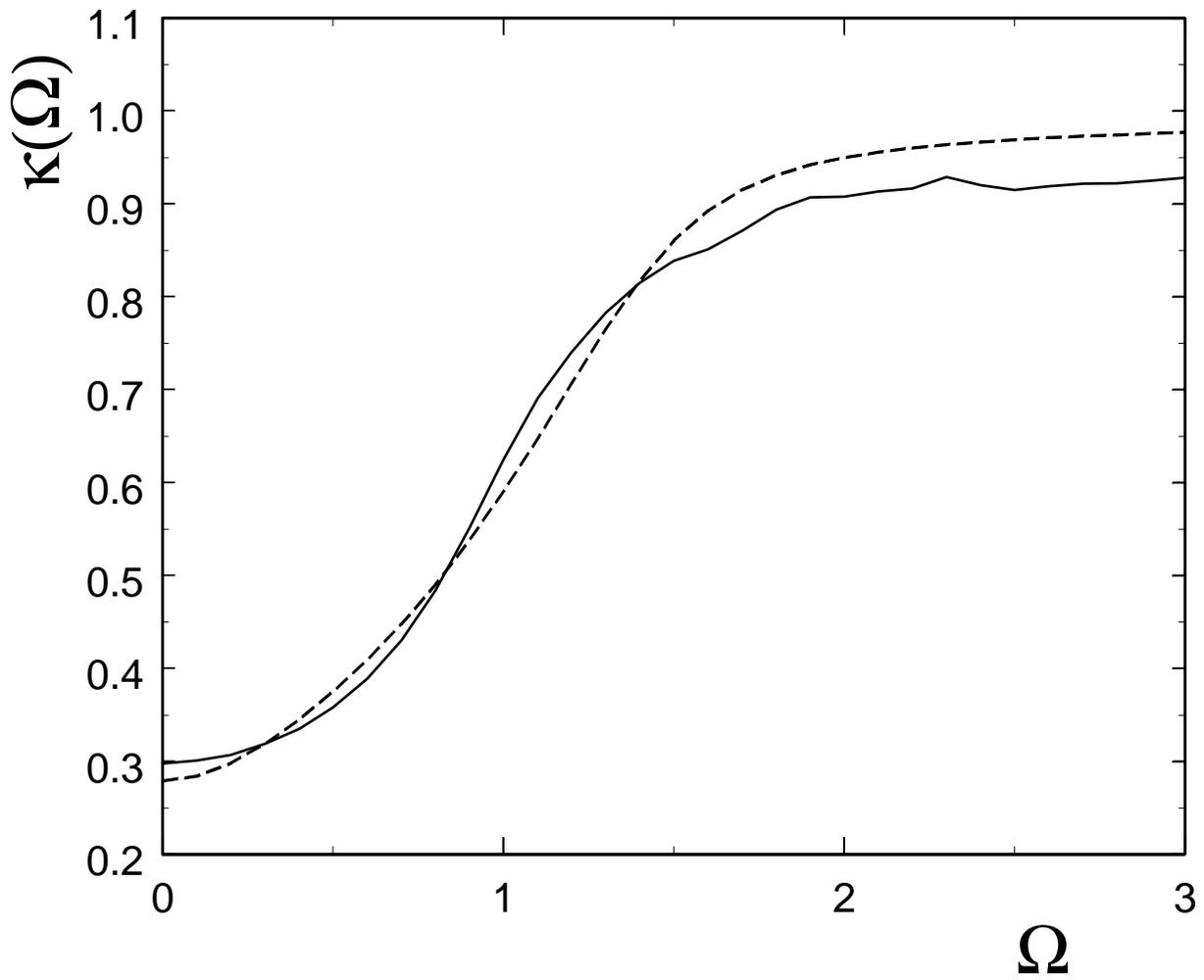
\end{document}